\documentclass[12pt,notitlepage]{article}
\usepackage{eucal,amsmath,amsfonts,amsthm,amssymb,latexsym,epsfig}

\theoremstyle{plain}

\newtheorem{thm}{Theorem}

\newtheorem{Lma}{Lemma}
\newtheorem{Crly}{Corollary}

\theoremstyle{definition}

\newcommand{\RR}{\mathbb{R}}

\newcommand{\rd}{\mathrm{d}}
\newcommand{\rw}{\mathrm{w}}

\newcommand{\lp}{\overline{\lim}}
\newcommand{\lf}{\underline{\lim}}

\begin{document}
\setcounter{page}{0}

\title{A Classification of Spherically Symmetric Static Solutions of SU(2) Einstein Yang Mills Equations with  
Non-negative Cosmological Constant}
\author{Alexander N. Linden\thanks{Zorn Visiting Assistant Professor, Indiana University}}
\date{}
\maketitle
\thispagestyle{empty}
%\startthechapters

% Input the individual chapters. 
%\unnumberedpage
%\setcounter{page}{1}
%\setcounter{figure}{0}

\begin{abstract}
\noindent
We categorize the global structure of spherically symmetric static solutions of Einstein SU(2)-Yang Mills equations with positive cosmological constant that are smooth at the origin.

\end{abstract}
\noindent
\section{Introduction}
Einstein's equations coupled to an SU(2) Yang Mills field have been the subject of numerous studies in the last decade.  Most have considered the system without cosmological constant and little has been proved rigorously when a positive cosmological constant $\Lambda$ is entered into the equations.  Recently, descriptions of solutions have been given by numerical analyses.  In this paper, we consider the coupled system with nonnegative cosmological constant and establish the global nature of solutions that are smooth at the origin of spherical symmetry. In particular, we establish that every solution gives rise to one of the five following geometries:\\ \\
%\textbf{Type I}: The spacelike hypersurface for each constant $t$ has the topology of a cylinder of constant radius.  These are Nariai solutions \cite{hN51}.\\ \\
\textbf{Type I}: The spacelike hypersurface for each constant $t$ has the topology of $S^3$.  One pole is the center of spherical symmetry and the opposite pole has a Reissner-Nordstr\"om like singularity.  There is also an horizon at the equator.  These are the generic solutions when $\Lambda=0$ \cite{pB94}.  We continue to call them \emph{generic} solutions.\\ \\
\textbf{Type II}: The spacelike hypersurface for each constant $t$ has the topology of $\RR^3$, there is an horizon at some $r_c\le \sqrt{3/\Lambda}$ and in the far field spacetime approaches Schwarzschild~-~ deSitter space, the space obtained with the metric
\begin{equation}\label{deSitter}
\rd s^2 = (1-\frac{2M}{r}-\frac{\Lambda r^2}{3}) \rd t^2 -(1-\frac{2M}{r}-\frac{\Lambda r^2}{3})^{-1}\rd r^2 - r^2 \rd \Omega^2.
\end{equation}
($\rd \Omega^2$ is the standard metric for the unit 2-sphere in spherical coordinates.)  We call these \emph{noncompact} solutions.  Their existence, for small $\Lambda$ is proved in \cite{L2}. \\ \\
\textbf{Type III}: The spacelike hypersurface for each constant $t$ has the topology of $\RR^3$.  Unlike the noncompact solutions, however, there is no horizon and the metric becomes singular in the far field.  Furthermore, a Yang-Mills connection term has an infinite number of zeros.  We call these solutions \emph{oscillating}.  
\\ \\
\textbf{Type IV}:  The spacelike hypersurface for each constant $t$ has the topology of $S^3$.   Both the metric and the Yang Mills connection are smooth everywhere.  We call such solutions \emph{compact regular} solutions.  An example is Einstein space.\\\\
\textbf{Type V}:  As with generic and compact regular solutions, the spacelike hypersurface for each constant $t$ has an horizon at a maximum $r_c$.  However, beyond the horizon, the solution does not continue back to the origin but rather to another horizon at some $r_0>0$.  As with noncompact solutions, this singularity can be transformed away by a Kruskal-like change of coordinates in which $r$ decreases through the singularity.  We call these solutions \emph{black hole} solutions. 

We prove the existence of solutions of each of these five types and also prove that every solution is one of these types. 

We begin with a spherically symmetric metric
\begin{equation}\label{metric}
\rd s^2=C^2A\;\rd t^2-\frac{1}{A}\;\rd r^2-r^2(\rd \phi^2+\sin^2\phi\;\rd\theta^2)
\end{equation}
and spherically symmetric Yang Mills connection
\begin{equation}\label{YMc}
a\mathsf{\sigma_3}\;\rd t+b\mathsf{\sigma_3}\;\rd r+\rw\mathsf{\sigma_2}\;\rd \phi+(\cos\phi\mathsf{\sigma_3}-\rw\sin\phi\mathsf{\sigma_1})\;\rd \theta.
\end{equation}
$\mathsf{\sigma}_i$ are the following matrices which form a basis of $su(2)$:
\[\mathsf{\sigma_1}=i/2\left[
\begin{array}{cc}
0 & -1\\
-1 & 0
\end{array}
\right],\;
\mathsf{\sigma_2}=i/2\left[
\begin{array}{cc}
0 & i\\
-i & 0
\end{array}
\right],\;
\mathsf{\sigma_3}=i/2\left[
\begin{array}{cc}
-1 & 0\\
 0 & 1
\end{array}
\right].
\]
With suitable choice of gauge, the assumption of a static magnetic field allows us to eliminate $a$ and $b$ from Equation~(\ref{YMc}).  
The Einstein and Yang Mills equations become a system of ordinary differential equations for $A$, $C$ and $\rw$ in the variable $r$.
\begin{equation}\label{Aeq}
rA'+2A{\rw'}^2=1-A-\frac{(1-\rw^2)^2}{r^2}-\Lambda r^2
\end{equation}
\begin{equation}\label{weq}
r^2A\rw''+r\rw'(1-A-\frac{(1-\rw^2)^2}{r^2}-\Lambda r^2)+\rw(1-\rw^2)=0
\end{equation} 
\begin{equation}\label{Ceq}
\frac{C'}{C}=\frac{2{\rw'}^2}{r}.
\end{equation}

Here, and throughout the paper $(')$ denotes a derivative with respect to $r$. 
To simplify notation, we define 
\begin{equation}\label{Phi}
\Phi(r)=1-A-\frac{(1-\rw^2)^2}{r^2}-\Lambda r^2.
\end{equation}
There exists a one-parameter family of solutions that are smooth at the origin; i.e., a family of solutions that satisfy
\begin{equation}\label{family}
(A(0),\rw(0),\rw'(0),\rw''(0))=(1,1,0,-\lambda),\;\;\;\lambda\ge 0
\end{equation}
(\cite{jS91}).  Because of the symmetry of Equations~(\ref{Aeq}) and (\ref{weq}) under the transformation $\rw\mapsto -\rw$, there is a complementary one-parameter family of solutions (namely those obtained by reflection) that satisfy 
\begin{equation}\label{familyref}
(A(0),\rw(0),\rw'(0),\rw''(0))=(1,-1,0,\lambda),\;\;\;\lambda\ge 0.
\end{equation}
Unless explicitly stated otherwise, throughout this paper any solution is assumed to be a member of one of these families.

In the case $\Lambda=0$, there are three possible types of solutions (i.e.,\cite{pB94}).  The first are \emph{particlelike} solutions \cite{jS93}.  These solutions are smooth for all $r>0$ and as $r\rightarrow \infty$, $(A,\rw^2,\rw')\rightarrow (1,1,0)$; i.e., in the far field, the metric becomes asymptotic to the Minkowski metric and the Yang-Mills field vanishes.

Also with $\Lambda=0$, there exist oscillating solutions \cite{pB94}.  For these solutions, $A(1)=0$ thus making Equations~(\ref{Aeq}) and~(\ref{weq}) singular at $r=1$.  As $r$ approaches $1$, $\rw$ approaches $0$ whereas $\rw'$ has no limit.  Moreover, $\rw$ has an infinite number of zeros \cite{pB94}.

The third type of solutions that exist when $\Lambda=0$ are generic solutions.  When $\Lambda=0$ these solutions are described as follows:\\
We define 
\[\Gamma=\{(r,A,\rw,\rw'):r>0,\; A>0,\;(\rw,\rw')\ne (0,0),\;\mathrm{and}\;\rw^2<1\}.\]  For generic solutions there exists an $r_e>0$ such that $(r,A(r),\rw(r),\rw'(r))\notin \Gamma$ whereas $A(r_e)>0$.  It follows from Equations~(\ref{Aeq}) and (\ref{weq}) that there exists an $r_c>r_e$ for which $A(r_c)=0$ and $\rw'(r_c)=\pm\infty$.  As a result, Equations~(\ref{Aeq}) and (\ref{weq}) become singular at this ``crash point''.  However, the solutions can be extended to type I solutions and $r$ attains a maximum value of $1$ \cite{pB94}.  Solutions of this type are generic in the sense that they remain of this type under small perturbations of $\Lambda$ and $\lambda$.

\section{Known Solutions}
\noindent
We begin by describing some explicit solutions, not necessarily smooth at the origin, when $\Lambda>0$.  Changing parameters from $r$ to $\rho$ by $C\dot r=1$ (dot ($\dot{\;})=\rd/\rd \rho$) transforms Equations~(\ref{Aeq}),~(\ref{weq}), and~(\ref{Ceq}) to the following:
\begin{equation}\label{Req}
r\ddot{r}=-2{\dot\rw}^2,
\end{equation}
\begin{equation}\label{Qwdoteq}
(Q\dot\rw)\dot{\;}=-\frac{\rw(1-\rw^2)}{r^2},
\end{equation}
and
\begin{equation}\label{Qreq}
\dot{Q}r\dot r+Q\dot r^2=2(Q\dot\rw^2-\frac{(1-\rw^2)^2}{2r^2})+1-\Lambda r^2
\end{equation}
where
\begin{equation}\label{Q}
Q=C^2A.
\end{equation}
The metric components are
\begin{equation}\label{rhometric}
\rd s^2 = Q \rd t^2 - \frac{1}{Q}\rd \rho^2 - r(\rho)(\rd \phi^2+\sin^2\phi\;\rd \theta^2)
\end{equation}
Differentiating Equation~(\ref{Qreq}) gives
\begin{equation}\label{Qddoteq}
\ddot Q=2Q(\frac{\dot r}{r})^2-\frac{2}{r^2}+4\frac{(1-\rw^2)^2}{r^4}.
\end{equation}
If $\dot r\equiv 0$ in a neighborhood of some $\rho_0$, then 
obviously $r$ is an unsuitable coordinate.  In this case Equations~(\ref{Req})~-~(\ref{Qreq}) yield
\begin{equation}\label{wNai}
\rw(1-\rw^2)\equiv 0.
\end{equation} 
If $\rw^2\equiv1$ then 
\begin{equation}\label{rNai1}
r(\rho)\equiv\frac{1}{\sqrt{\Lambda}}
\end{equation}
and 
\begin{equation}\label{QNai1}
Q(\rho)=-\Lambda \rho^2+a\rho+b.
\end{equation}

If $\rw\equiv 0$, then
\begin{equation}\label{rNai}
r(\rho)\equiv\sqrt{\frac{1\pm\sqrt{1-4\Lambda}}{2\Lambda}}
\end{equation}
and 
\begin{equation}\label{QNai0}
Q(\rho)=\frac{2}{r^2}(\frac{2}{r^2}-1)\rho^2+a\rho+b.
\end{equation}
(Here $a$ and $b$ are constants of integration.)  These are $H^2\times S^2$ Nariai solutions \cite{hN51}. 

If $\dot r\not\equiv 0$ in a neighborhood of $\rho_0$, it follows from Equation~(\ref{Ceq}) that $C<\infty$ in a neighborhood of $\rho_0$, and consequently, that $\rho_0$ is an isolated zero of $\dot r$.  Thus, we may take $r$ as a coordinate for $r<r(\rho_0)$ and the geometry of spacetime can be determined by solutions of Equations~(\ref{Aeq})~-~(\ref{Ceq}).  Equation~(\ref{Ceq}) can be ignored because it separates from the other two. 

For any positive $\Lambda$, there exists an $r_c\in(0,\sqrt{3/\Lambda})$ such that $A(r_c)=0$ at which point Equations~(\ref{Aeq}) and (\ref{weq}) become singular \cite{L2}.  There also exist solutions such that $\rw^2(r_c)\le 1$ and ${\rw'}^2(r_c)<\infty$.  An example of this type of solution is the following:
\begin{equation}\label{RN}
A=1-\frac{\Lambda r^2}{3},\;\;\;\;\rw\equiv 1
\end{equation}
which gives the metric of Schwarzschild~-~deSitter space. If $\Lambda$ is small, there exist other solutions for which $\rw^2(r_c)<1$ and ${\rw'}^2(r_c)<\infty$ \cite{L2}. For all such solutions, the singularity is only a coordinate singularity and $r$ increases in the extended solution.  Moreover, the extended solution is smooth for all $r>r_c$ and the geometry approaches that of deSitter space \cite{L3}.  These are the noncompact solutions of Type II.

Another explicit solution is Einstein space which can described as follows: $\Lambda=3/4$, $A(r)=1-r^2/2$ and  $\rw=\pm \sqrt{A}$.  The spacelike hypersurface has the topology of $S^3$ with maximum radius $r_c=\sqrt{2}$.  $A$ is symmetric  and $\rw$ is antisymmetric with respect to a refection through the equatorial 3 dimensional hyperplane.  This is a compact regular solution.  Under a change of coordinates this solution can be described globally by
\begin{equation}\label{Ein}
\rd s^2=\rd t^2-\frac{8e^{2\tau}}{(1+e^{2\tau})^2}(\rd \tau^2+\rd \Omega^2).
\end{equation}
\begin{equation}\label{wein}
\rw=\frac{1-e^{2\tau}}{1+e^{2\tau}}.
\end{equation}

Although not smooth at $r=0$, there are other solutions of interest.  Among these are the Reissner-Nordtr\"{o}m solutions:
\begin{equation}\label{RNg}
A=1-\frac{c}{r}-\frac{\Lambda r^2}{3}\;\;\;\rw\equiv 1
\end{equation}
and
\begin{equation}\label{ERN}
A=1-\frac{c}{r}+\frac{1}{r^2}-\frac{\Lambda r^2}{3}\;\;\;\rw\equiv 0
\end{equation} 
where $c$ is an arbitrary constant.  A special case of Equation~(\ref{RNg}) is when $c=0$ in which case Equation~(\ref{RNg}) becomes Equation~(\ref{RN}).  If for any $\tilde r$, $A(\tilde r)>0$ and $\rw(1-\rw^2)(\tilde r)=\rw'(\tilde r)=0$, then Equations~(\ref{Aeq}) and~(\ref{weq}) are smooth near $\tilde r$ and standard uniqueness theorems imply that the solution must be either that of Equation~(\ref{RNg}) or Equation~(\ref{ERN}), depending on the value of $\rw(\tilde r)$.

 What remains is to establish the existence of oscillating and black hole solutions and to prove that every solution is one of the five aforementioned types.

\section{Behavior near $r_c$}
 The global geometry of solutions depends on the behavior of $\rw'$ at $r_c$.  We analyze  it by studying the local behavior of $\rw$ and $A$ near $r_c$.  Much of our analysis  addresses the question of limits of variables as $r\nearrow r_c$.  To simplify notation,  once a limit for a variable is established, we denote this limit with the subscript ``c''.

Because a solution is a noncompact solution whenever $\lp_{r\nearrow r_c}{\rw'}^2(r)<\infty$, we need consider only the case $\lp_{r\nearrow r_c}\rw'(r)=\infty$.  We take advantage of the symmetry of Equation~(\ref{Aeq}) and~(\ref{weq}) and consider only the case $\lp_{r\nearrow r_c}\rw'(r)=+\infty$. 
\begin{Lma}\label{A=0}
$\lim_{r\nearrow r_c}A(r)=0$.
\end{Lma}
\noindent
\textbf{Proof}:  $r_c$ is defined by $\lf_{r\nearrow r_c}A(r)=0$.  The result follows immediately from Equation~(\ref{Aeq}).\hfill $\blacksquare$  
\begin{Lma}\label{w>1w'inf}
If there exist $r<r_c$ such that $\rw(r_c)>1$, then $\lim_{r\nearrow r_c}\rw'(r)=+\infty$.
\end{Lma}
\noindent
\textbf{Proof}:  If there exists an $\bar r<r_c$ such that $\rw(\bar r)>1$ then such an $\bar r$ can be found such that $\rw'(\bar r)>0$ also.  It follows easily from Equation~(\ref{weq}) that $\rw\ge 0$ for all $r>\bar r$.  For any $M>0$, if there exists a sequence $\{r_n\}\nearrow r_c$ such that $\rw'(r_n)=M$, then Equation~(\ref{weq}) implies that for $n$ sufficiently large, $\rw''(r_n)>0$.  It follows that $\rw'$ has a limit $\rw'_c$ as $r\nearrow r_c$.  We now consider the following equation which is obtained easily from Equations~(\ref{Aeq}) and~(\ref{weq}):
\begin{equation}\label{veq}
r^2(A\rw')'+2r{\rw'}^2(A\rw')+\rw(1-\rw^2)=0.
\end{equation}
If $\rw'_c<\infty$, then, since $A(r_c)=0$, $\lim_{r\nearrow r_c}A\rw'(r)=0$.  However, it is clear from Equation~(\ref{veq}) that this is impossible. \hfill $\blacksquare$
\begin{Lma}\label{w<infty}
$\lp_{r\nearrow r_c}\rw(r)<\infty$.
\end{Lma}
\noindent
\textbf{Proof}:  In light of Lemma~\ref{w>1w'inf}, without any loss of generality, we assume that $\lim_{r\nearrow r_c}\rw(r)=+\infty$ and arrive at a contradiction. 
Equation~(\ref{Phi}) implies readily that 
\begin{equation}\label{phiasym}
\Phi(r)\sim-\rw^4
\end{equation}
Equation~(\ref{weq}) gives 
\begin{equation}\label{w''asym}
r^3A\rw''\sim\rw^4\rw'
\end{equation}
as $r\nearrow r_c$.  Also, from Equation~(\ref{w''asym}) and the fact that $A>0$ for $0<r<r_c$, it follows that 
\begin{equation}\label{w''inf}
\rw''(r)=\infty.
\end{equation}
Our assumptions and Equation~(\ref{w''inf}) now permit the application of L'H\^opital's rule which, together with Equation~(\ref{w''asym}), yields
\begin{equation}\label{hop}
\lim_{r\nearrow r_c}\frac{\rw^5}{\rw'}=\lim_{r\nearrow r_c}\frac{5r^3A\rw^4\rw'}{r^3A\rw''}=\lim_{r\nearrow r_c}\frac{5r^3A\rw^4\rw'}{\rw^4\rw'}=0.
\end{equation}
We now define the following important variable:
\begin{equation}\label{z}
z(r)=\frac{\Phi(r)+2A{\rw'}^2(r)}{r}.
\end{equation}
A simple calculation using Equations~(\ref{Aeq}) and (\ref{weq}) gives
\begin{equation}\label{zeq}
r^2z'+2r{\rw'}^2z+2(1-A-\frac{2(1-\rw^2)^2}{r^2})=0.
\end{equation}
The usefulness of $z$ lies in knowledge of the sign of the last term on the left side of Equation~(\ref{zeq}).  This enables the establishment of limits for all the important variables as $r$ approaches $r_c$.

We assert that $\lim_{r\nearrow r_c}z(r)$ exists and that
\begin{equation}\label{zlim}
-\infty<\lim_{r\nearrow r_c}z(r)<\infty.
\end{equation}
Indeed, Equation~(\ref{zeq}) implies that $z'(r)>0$ whenever $z(r)=0$ and $r$ is sufficiently close to $r_c$ so that $2(1-\rw^2)^2>r^2$.  From this it follows that $z$ can have only one sign near $r_c$.  If $z<0$ near $r_c$, then (\ref{zlim}) holds because $z$ is always negative and increasing.  If $z>0$ near $r_c$, then we divide Equation~(\ref{zeq}) by $\rw'$ and invoke Equation~(\ref{hop}).  This gives, for any $\epsilon>0$, any sequence $\{r_n\}\nearrow r_c$ that satisfies $z(r_n)>\epsilon$ must also satisfy $z'(r_n)<0$.  Inequalities~(\ref{zlim}) follow. 

We now consider the following equation for any constant $\beta>0$:
\begin{equation}\label{aw'beta}
r^2(A{\rw'}^\beta)'+r{\rw'}^{\beta}[rz+(\beta-2)\Phi+\frac{\beta\rw(1-\rw^2)}{\rw'}]=0.
\end{equation}
Equation~(\ref{aw'beta}) follows easily from Equations~(\ref{Aeq}) and (\ref{weq}).  Equations~(\ref{phiasym}), (\ref{hop}), (\ref{zlim}), and (\ref{aw'beta}) yield, for any $\beta\in(0,2)$, $\lim_{r\nearrow r_c}(A{\rw'}^{\beta})'=-\infty$; i.e., $\lim_{r\nearrow r_c}A{\rw'}^\beta$ exists and is finite.  By choosing arbitrary $\bar\beta\in (\beta,2)$, we conclude, on one hand, that 
\begin{equation}\label{aw'b=0}
\lim_{r\nearrow r_c}A{\rw'}^\beta=\lim_{r\nearrow r_c}(A{\rw'}^{\bar\beta})({\rw'}^{\beta-\bar\beta})=0\;\;\;\mathrm{for\;all}\;\beta\in(0,2).
\end{equation}

On the other hand, we assert that (\ref{aw'b=0}) cannot hold.  Indeed, Equation~(\ref{Aeq}) implies that $\lim_{r\nearrow r_c}A'(r)=-\infty$. Consequently, there exist $\delta>0$ such that 
\begin{equation}\label{A>rho}
A(r)>(r_c-r)\;\;\;\mathrm{whenever\;}r\in(r_c-\delta,r_c).
\end{equation}
Taking $\delta$ to be smaller, if necessary, yields
\begin{equation}\label{w'betasmall}
{\rw'}^{\beta}=\frac{A{\rw'}^{\beta}}{A}<\frac{1}{(r_c-r)}\;\;\;\mathrm{whenever}\;r\in(r_c-\delta,r_c);
\end{equation}
 i.e.,
\begin{equation}\label{w'small}
\rw'<\frac{1}{(r_c-r)^{1/\beta}}\;\;\;\mathrm{whenever}\;r\in(r_c-\delta,r_c);
\end{equation}
Integrating inequality~(\ref{w'small}) from any $\bar r\in(r_c-\delta,r_c)$ to $r_c$ implies that $\lim_{r\nearrow r_c}\rw(r)<\infty$.  However, we have assumed that $\lim_{r\nearrow r_c}\rw(r)=\infty$.  This completes the proof. \hfill $\blacksquare$\\

We will have frequent occasion to use Equation~(\ref{aw'beta}) with $\beta=2$.  Hence, we define the variable
\begin{equation}\label{f}
f=A{\rw'}^2
\end{equation}
and write, as a separate equation, the following:
\begin{equation}\label{feq}
r^2f'+r^2{\rw'}^2z+2\rw\rw'(1-\rw^2)=0.
\end{equation}
\begin{Lma}\label{phinp}
For all solutions of Equations~(\ref{Aeq}) and~(\ref{weq}), noncompact or not, $\lf_{r\nearrow r_c}\Phi(r_c)\le 0$.
\end{Lma}
\noindent
\textbf{Proof}: If $\lf_{r\nearrow r_c}\Phi(r)>0$, then Equation~(\ref{weq}) and Lemma~\ref{w<infty} imply that $\lp_{r\nearrow r_c}{\rw'}^2(r)<\infty$.  Consequently $\lim_{r\nearrow r_c}\Phi(r)$ exists and is positive.  Moreover, $\lim_{r\nearrow r_c}A{\rw'}^2(r)=0$.  Equation~(\ref{Aeq}) now gives $\lim_{r\nearrow r_c}A'(r)>0$ which is impossible.\hfill $\blacksquare$\\

Lemmas~\ref{w<infty} and \ref{phinp} enable us to establish limits of $A'$, $f$ and $\Phi$ as $r\rightarrow r_c$.  A priori, there are four possibilities.  One is that $A'$ or $\Phi$ is without limit as $r\nearrow r_c$.  A second possibility is that $\lp_{r\nearrow r_c}{\rw'}^2(r)<\infty$.  Thirdly, there could be solutions for which $\lf_{r\nearrow r_c}\rw'(r)<\lp_{r\nearrow r_c}\rw'(r)$.  Finally, $\lim_{r\nearrow r_c}{\rw'}(r)$ could be infinite.  We will prove that there are no solutions for which $A'$ is without limit whereas, provided $\Lambda$ is small,  each of the other possibilities is satisfied for some solution.

\section{$2\Lambda r_c^2< 1$}
Solutions in which $2\Lambda r_c^2<1$ are relatively simple to analyze.  As we shall see, this is because we know the sign of $z$ near $r_c$.  As $2\Lambda r_c^2<1$ always whenever $\Lambda=0$, the complications added by considering nonzero $\Lambda$ lie in the possibility that $2\Lambda r_c^2>1$.

Throughout this section, we assume $2\Lambda r_c^2<1$.

\begin{Lma}\label{limits}
$f$, $\Phi$, and $A'$ have finite limits as $r\rightarrow r_c$.  Moreover, $z\not\equiv 0$ in a neighborhood of $r_c$.
\end{Lma}
\noindent
\textbf{Proof}:  We prove that $1-A-2(1-\rw^2)^2/r^2$ has the same sign on any sequence $\{r_n\}$ approaching $r_c$ such that $z(r_n)=0$.  From this and Equation~(\ref{zeq}) it follows that $z$ has only one sign near $r_c$.  Equation~(\ref{zeq}) then yields a bound on $z'$ which implies a finite limit for $z$ as $r\nearrow r_c$. 

Clearly, whenever $z(r)=0$, $\Phi(r)\le 0$.  Also, because $2\Lambda r_c^2<1$, there exist $\epsilon>0$ such that whenever $\Phi(r)\le 0$, 
\[\frac{(1-\rw^2)^2}{r^2}\ge 1-A-\Lambda r^2>\frac{1}{2}+\epsilon-A.\]
Consequently,
\[1-A-2\frac{(1-\rw^2)^2}{r^2}<A-2\epsilon\]
i.e.,
\[1-A-2\frac{(1-\rw^2)^2}{r^2}<0\]
whenever $\Phi\le 0$ and $r$ is sufficiently close to $r_c$ so that $A(r)<\epsilon$.  This proves that $z$ has a limit $z_c$ as $r\nearrow r_c$ and also that $z\not\equiv 0$ near $r_c$.

We will prove shortly that $\lim_{r\nearrow r_c}z(r)=0$.  However, we need not assume this to prove that $f$ has a finite limit.   If $z_c>0$, then for any $\epsilon>0$, and any $r$ sufficiently close to $r_c$ and such that $f(r)=\epsilon$, ${\rw'}^2(r)$ is large and, consequently, Equation~(\ref{feq}) implies that $f'(r)<0$.  It follows that $\lim_{r\nearrow r_c}f(r)$ exists.  A similar argument proves that $f$ has a limit whenever $z_c<0$.  Since $z$ and $\Phi$ are bounded, this limit must be finite.  Finally, if $z_c=0$, then $f$ has a finite limit \cite{L2}.  \hfill $\blacksquare$\\[10cm]
\begin{thm}\label{smbag}
Whenever $\Phi_c<0$, $z_c=0$.  Moreover, as $r\nearrow r_c$, $z$ approaches zero from above.
\end{thm}
\noindent
\textbf{Proof}: Similarly as in \cite{pB94}, we introduce the new variables
\begin{equation}\label{N}
N=\sqrt{A},
\end{equation}
\begin{equation}\label{U}
U=N\rw'
\end{equation}
\begin{equation}\label{k}
\kappa=\frac{1}{2N}(rz+2N^2),
\end{equation}
and the a new  parameter $\tau$ defined by $\rd r/\rd \tau = rN$.  Equations~(\ref{Aeq}),~(\ref{weq}) and~(\ref{Ceq}) transform to
\begin{equation}\label{req}
\dot r = rN,
\end{equation}
\begin{equation}\label{wtaueq}
\dot\rw = rU,
\end{equation}
\begin{equation}\label{Neq}
\dot N = (\kappa-N)N - 2U^2,
\end{equation}
\begin{equation}\label{Ueq}
\dot U = -\frac{\rw(1-\rw^2)}{r}-(\kappa-N)U,
\end{equation}
\begin{equation}\label{Ctau}
\dot{CN}=(\kappa-N)CN
\end{equation}
where dot $(\dot{\;})$ denotes $\rd/\rd \tau$.  We also have the auxilliary equation
\begin{equation}\label{keq}
\dot \kappa = 1 + 2U^2-\kappa^2-2\Lambda r^2.
\end{equation}
The metric transforms to 
\begin{equation}\label{taumet}
\rd s^2 = C^2 N^2\rd t^2 - r^2 (\tau)(\rd \tau^2 + \rd \Omega^2).
\end{equation}
Because $\Phi_c<0$, $A'_c<0$.  Equation~(\ref{Aeq}), its integral, Equation~(\ref{weq}), and its derivatives yield the following limits:
\begin{eqnarray}\label{Nlims}
\lim_{r\nearrow r_c}A'(r)&=&A'_c<0\nonumber\\
\lim_{r\nearrow r_c}\frac{A(r)}{r_c-r}&=&A'_c\nonumber\\
\end{eqnarray}

 It follows from these limits and integrating Equation~(\ref{req}) that there exists a finite $\tau_c$ such that $r_c=r(\tau_c)$ and that near $\tau_c$, $(r_c-r)\sim (\tau_c-\tau)^2$.  Equation~(\ref{keq}) gives $\kappa>0$ for all $\tau>-\infty$ and $\lp_{\tau_\nearrow \tau_c}\kappa(\tau)<\infty$.  Consequently, the set of Equations~(\ref{req}) -~(\ref{keq}) and the metric~(\ref{taumet}) are nonsingular at $\tau_c$.  The result now follows readily from solving Equation~(\ref{k}) for $z$ near $\tau_c$.  \hfill $\blacksquare$\\

With $\tau$ as the parameter of equations~(\ref{req})~-~(\ref{keq}) we define $\tau_c$ to be the smallest $\tau>-\infty$ such that
\begin{equation}\label{tauc}
r_c(\tau)=\tau_c
\end{equation}
and conclude the following:
\begin{Crly}\label{tcfin}
If $\Phi_c<0$, then $\tau_c<\infty$ and Equations~(\ref{req})~-~(\ref{keq}) as well as the metric~(\ref{taumet}) are nonsingular at $\tau_c$.
\end{Crly}
\begin{Crly}\label{bog}
If $\Phi_c<0$, then the solution $(A,\rw)$ is generic.
\end{Crly}
\noindent
\textbf{Proof}:  The solution to Equations~(\ref{req})~-~(\ref{Neq}) is valid up to some $\tau_0$ which is infinite unless $r$ approaches zero at least on a sequence $\{\tau_n\}\nearrow \tau_0$ or some other dependent variable becomes singular.  Theorem~\ref{smbag} and Equation~(\ref{Neq}) imply that $N<0$ for $\tau>\tau_c$.  In this region, we return to $r$ as our parameter and extend our original solution to a solution valid in the region $(r_0,r_c)$ where $r_0=r(\tau_0)$ is zero unless the solution is a black hole.  Since $\kappa>0$ and is bounded away from zero, Equation~(\ref{Neq}) excludes the possibility of a black hole.  The solution cannot be smooth at $r=0$ either.  Indeed, if it were, because $z\searrow 0$ as $r\nearrow r_c$ and $N<0$, $\kappa$ would be negative for some $\tau$ near $\tau_c$.  Clearly, this is impossible.\hfill $\blacksquare$

$\;$\\[10cm]
\begin{thm}\label{crit}
Whenever $\Phi_c=0$, either the solution is generic or $\rw_c=0$ and the solution is oscillating.
\end{thm}
\noindent
\textbf{Proof}:  $\kappa_c$ finite implies $z_c=0$.  Consequently, $A'_c=f_c=0$. $2\Lambda r_c^2<1$ and Equation~(\ref{Phi}) give $\rw_c^2\ne 1$.  We first prove that in the case $\rw_c\ne 0$, $\tau_c<\infty$.

Equation~(\ref{weq}) implies that $\rw'$ has only one sign near $r_c$.  Equation~(\ref{Ueq}) yields
\begin{equation}\label{u}
\rw_c\rw'_c(1-\rw_c^2)>0.
\end{equation} 
Without loss of generality we take $\rw'$ to be positive near $r_c$ and from Equation~(\ref{veq}) conclude that, near $r_c$,
\begin{equation}\label{Awbyw}
A=\frac{A\rw'}{\rw'}>\frac{\rw_c(1-\rw_c^2)(r_c-r)}{2r_c^2\rw'}.
\end{equation}
Since $A'_c=0$, it follows that $\lim_{r\nearrow r_c}\rw'(r)=\infty$.

Differentiating Equation~(\ref{Aeq}) gives
\begin{equation}\label{A''}
A''=\rw'[\frac{-2A'}{r\rw'}+\frac{2\rw'z}{r}+\frac{2(1-\rw^2)^2}{r^4\rw'}-\frac{2\Lambda}{\rw'}+\frac{8\rw(1-\rw^2)}{r^3}].
\end{equation}
Multiplying and dividing Equation~(\ref{k}) by $\rw'$ gives 
\begin{equation}\label{w'z0}
\lim_{r\nearrow r_c}\rw'z(r)=0
\end{equation}
 from which it is clear that
\begin{equation}\label{A''asym}
A''(r)\sim \frac{8\rw\rw'(1-\rw^2)}{r^3}
\end{equation}
as $r\nearrow r_c$. 

Differentiating Equation~(\ref{weq}) yields $\rw'''(r)<0$ for any $r$ near $r_c$ such that $\rw''(r)=0$.    Therefore, 
\begin{equation}\label{w''>0}
\rw''(r)\ge 0
\end{equation}
for all $r$ near $r_c$.
Equations~(\ref{weq}), (\ref{w'z0}) and~(\ref{w''>0}) yield the existence of positive $\eta$ such that
\begin{equation}\label{aw'3big}
\lf_{r\nearrow r_c}A{\rw'}^3(r)>\eta.
\end{equation}
Consequently,
\begin{equation}\label{w'3}
{\rw'}^3=\frac{A{\rw'}^3}{A}>\frac{\eta}{\epsilon(r_c-r)}
\end{equation}
for arbitrarily small positive $\epsilon$ provided $r$ is sufficiently close to $r_c$.  It follows from Equations~(\ref{A''asym}) and~(\ref{w'3}) that
\begin{equation}\label{A''bound}
A''>\frac{1}{(r_c-r)^{1/3}}
\end{equation}
near $r_c$.  Integrating Equation~(\ref{A''bound}) twice gives
\begin{equation}\label{Nbound}
\sqrt{A}>\frac{3}{\sqrt{10}}(r_c-r)^{5/6}
\end{equation}
near $r_c$.
 From Equation~(\ref{req}) it now follows that $\tau_c<\infty$. Thus, the solution ($r,N,U,C,k)$ can be continued beyond $\tau_c$. Moreover, $N(\tau_c)=\dot N(\tau_c)=\ddot N(\tau_c)=0$ whereas  $N\dot{\;}\dot{\;}\dot{\;}(\tau_c)=-4\rw_c^2(1-\rw_c^2)^2/r^2_c<0$.  Therefore,  $N<0$ for $\tau$ near to but greater than $\tau_c$.  $\kappa_c$ positive and finite gives $z\sim(\tau_c-\tau)^3$ with $z$ approaching zero from above as $\tau\nearrow \tau_c$.  As in Corollary~\ref{bog}, the solution must be generic.  

The other possibility is that $\rw_c=0$.  To prove that such a solution is oscillating we note that all derivatives of Equations~(\ref{Neq}),~(\ref{wtaueq}), or~(\ref{Ueq}) vanish at $\tau_c$.  This can only be if $\tau_c=\infty$.  Also, $\dot \kappa=0$.  We now define
\begin{equation}\label{Theta}
\Theta(\tau)=\mathrm{Arctan}\frac{\dot\rw}{\rw}.
\end{equation}
For any solution of Equations~(\ref{req})~-~(\ref{keq}), $\Theta$ must satisfy the following equation:
\begin{equation}\label{Theq}
\dot\Theta=-1+\rw^2\cos^2\Theta+(2N-\kappa)\sin\Theta\cos\Theta
\end{equation}
where
\[\cos\Theta=\frac{\rw}{\sqrt{\rw^2+r^2U^2}}\]
and
\[\sin\Theta=\frac{rU}{\sqrt{\rw^2+r^2U^2}}.\]
$\dot\kappa\rightarrow 0$ implies that $\kappa\rightarrow 1-2\Lambda r_c^2<1$.  It follows from Equation~(\ref{Theq}) and the fact that $\tau_c=\infty$ that the solution is oscillating.\hfill $\blacksquare$\\ \\
\subsection{Oscillating Solutions}
We now prove the existence of generic and oscillating solutions for $\Lambda<1/4$.   We begin by defining, as in \cite{jS91}
\begin{equation}\label{a-w2}
h(r)=A(r)-\rw^2(r)
\end{equation}
and
\begin{equation}\label{c=2}
g(r)=2r^2-(1-\rw^2)(r).
\end{equation}
\begin{Lma}[\cite{jS91}]\label{hneg}
For any $\Lambda\ge 0$, $\lambda>2$, $h(r)<0$ as long as $g(r)<0$.
\end{Lma}
\noindent
\textbf{Proof}:  Simple calculations give $h(0)=0=h'(0)=0$, and $h''(0)=2\lambda(1-\lambda)-2\Lambda/3$; hence
\[h(r)<0\;\;\;\mathrm{for}\;r\;\mathrm{near}\;0,\;\mathrm{if}\;\lambda>1.\]
Moreover, $g(0)=g'(0)=0$ and $g''(0)=(2-\lambda)$; thus
\[g(r)<0\;\;\;\mathrm{for}\;r\;\mathrm{near}\; 0,\;\mathrm{if}\;\lambda>2.\]
Whenever $\lambda>1$, if $\tilde r$ is the smallest $r>0$ such that $h(\tilde r)=0$, 
\begin{equation}\label{h'tilder}
h'(\tilde r)=\frac{\Phi}{r}-\frac{-2(\rw{\rw'})^2}{r}-2\rw\rw'.
\end{equation}
Considering the right side of Equation~(\ref{h'tilder}) as a quadratic form in $s=(\rw\rw')$ gives discriminant at $\tilde r$ 
\begin{eqnarray}\label{discr}
\Delta & = & 4(1+\frac{2\Phi}{r^2})=4[1+\frac{2(1-A)}{r^2}-\frac{2(1-\rw^2)^2}{r^4}-2\Lambda]\nonumber\\
 & = & 4[1+\frac{2(1-\rw^2)}{r^2}-\frac{2(1-\rw^2)^2}{r^4}-2\Lambda]\nonumber\\
 & = & 4[1+\frac{2(1-\rw^2)}{r^2}(1-\frac{1-\rw^2}{r^2})-2\Lambda]\nonumber\\
 & < & 4[1-\frac{2(1-\rw^2)}{r^2}-2\Lambda]<4(1-4-2\Lambda)<0
\end{eqnarray}
whenever $g(\tilde r)<0$.  Since $-(2/\tilde r) s^2-2s+\Phi/\tilde r<0$ for large negative $s$, it follows that $h(r)<0$ as long as $g(r)<0$.\hfill $\blacksquare$
\begin{Lma}[\cite{jS91}]\label{g'neg}
For any $\Lambda\ge 0$, $\lambda>2$, $g'(r)<0$ as long as $h(r)<0$.
\end{Lma}
\noindent
\textbf{Proof}:  

For any solution $(A,\rw)$ of Equations~(\ref{Aeq}) and~(\ref{weq}) $g$ satisfies the following equation:
\begin{equation}\label{geq}
g''(r)=-\frac{\Phi}{rA}g'+\frac{2}{r^2A}[r^2A{\rw'}^2-(1-\rw^2)^2-2\Lambda r^4+g].
\end{equation}
If $\lambda>2$ and $\tilde r$ is the smallest $r$ such that $g'(r)=0$,
\begin{eqnarray}\label{gtilder}
g''(\tilde r)&=&\frac{2}{\tilde r^2A}[r^2A{\rw'}^2-(1-\rw^2)^2-2\Lambda r^4+g]_{r=\tilde r}\nonumber\\
 &\le&\frac{2}{\tilde r^2A}[r^2\rw^2{\rw'}^2-(1-\rw^2)^2-4\Lambda r^4+g]_{r=\tilde r}\nonumber\\
&=&\frac{2}{\tilde r^2A}[4\tilde r^4-(1-\rw^2)^2-4\Lambda r^4+g]_{r=\tilde r}\nonumber\\
 &=&\frac{2}{\tilde r^2A}[g(1+2\tilde r^2+(1-\rw^2))-4\Lambda r^4]_{r=\tilde r}<0
\end{eqnarray}
whenever $h(\tilde r)$ and $g(\tilde r)$ are negative.  The result follows.\hfill $\blacksquare$\\[10cm]  
\begin{thm}[\cite{jS91}]\label{w>0}
For any $\Lambda\ge 0$, $\lambda>2$,  $\rw(r)>0$ for all $r\in[0,r_c)$.
\end{thm}
\noindent
\textbf{Proof}:  Since $h$ and $g$ are both negative near $0$, if there exists a $\hat r\in(0,r_c)$ such that $\rw(\hat r)=0$, then $h(\hat r)>0$ and, consequently, there exists a smallest $r$, say $r_1$ such that $h(r_1)=0$.  Similarly, $g'(\hat r)>0$ and there exists a smallest $r$, say $r_2$ such that $g'(r_2)=0$.  Now, from what has just been shown, both $r_1<r_2$ and $r_1>r_2$ must hold, which is clearly impossible.  The result follows.\hfill $\blacksquare$
\begin{Crly}\label{rc<1}
For any $\Lambda\ge 0$, $r_c<1/\sqrt{2}$ whenever $\lambda>2$.
\end{Crly}
\noindent
\textbf{Proof}:  If $r_c\ge 1/\sqrt{2}$, then $g(r_c)>0$.  It follows that there exists a smallest $r_2$ such that $g'(r_2)=0$.  Lemma~\ref{hneg} gives $h(r)<0$ for all $r<r_2+\delta_1$ for some small positive $\delta_1$.  Lemma~\ref{g'neg} then gives $g'(r)<0$ for all $r<r_2+\delta_2$ for some small positive $\delta_2$, contradicting the fact that $g'(r_2)=0$.  The result follows. \hfill $\blacksquare$  
\begin{thm}\label{oscexist}
Generic and oscillating solutions exist for $\Lambda<1/4$.
\end{thm}
\noindent
\textbf{Proof}:  For any generic solution, at $\tau_c$, $N$ changes sign and Equations~(\ref{req})~-~(\ref{keq}) are nonsingular. Continuous dependence on parameters implies that under small perturbations of $\Lambda$ or $\lambda$, solutions remain generic.  Also, for any fixed $\Lambda$, both $r_c$ and $\rw_c$ are continuous functions of $\lambda$.

We now define $r_1$ and $r_2$ to be the two positive roots of
\[1-\frac{1}{r^2}-\Lambda r^2\]
and take $r_1<r_2$.  

From Theorem~\ref{w>0}, for any $\lambda>2$, the solution $(A(\lambda,r),\rw(\lambda,r))$ crashes before $\rw=0$.  Corollary~\ref{rc<1} gives $r_c<1$ for any such solution.  Equation~(\ref{weq}) implies $\rw'<0$ for all $r\in(0,r_c)$; i.e., $U(r)<0$.  Also, from Equation~(\ref{keq}), $\kappa>0$ for all $\tau<\tau_c$ from which it follows that near $\tau_c$, $\dot U<0$.  Consequently, $f_c>0$ and such solutions are generic; i.e., for any $\Lambda<1/4$, and $\lambda_0>2$, $(A(\Lambda,\lambda_0,r),\rw(\Lambda,\lambda_0,r))$ is a generic solution and satisfies $r_c<1$ and $0<\rw_c<1$.  

To establish the existence of oscillating solutions, we fix $\Lambda$ and $\lambda_0$.  If possible, we choose $\lambda^2$ to be the largest $\lambda<\lambda_0$ such that $r_c(\lambda^2)=\sqrt{2}$ and $\lambda_1$ to be the largest $\lambda<\lambda_0$ such that $\rw^2_c(\lambda_1)=1$.  $\Phi_c(\lambda_1)\le 0$ and $\Lambda<1/2$ give $r_c(\lambda_1)>\sqrt{2}$.  If $(A(\lambda,r),\rw(\lambda,r))$ is generic for each $\lambda\in[\lambda_1,\lambda_0]$, then, because $r_c$ and $\rw_c$ are continuous functions of $\lambda$, $\lambda^2>\lambda_1$ and $\rw^2_c(\lambda_2)\le 1$.   However, this implies that $r_c(\lambda^2)<r_1<\sqrt{2}$ which contradicts the definition of $\lambda^2$.  Thus, there are three possibilities. One is that there is no $\lambda_1$.  Another is that there is no $\lambda^2$.  The third is that both $\lambda_1$ and $\lambda^2$ exist but there is a discontinuity in $r_c$ for some $\lambda_c$.  All three possibilities lead to the same conclusion; namely, there exists a $\lambda_c$ such that  $r_c(\lambda_c)<\sqrt{2}$ and $r_c$ is discontinuous at $\lambda_c$.  The solution $(A(\lambda_c,r),\rw(\lambda_c,r))$ must be oscillating.\hfill $\blacksquare$\\

We conclude this section with the following remarks about oscillating solutions:\\ \\
(1) The proof of Theorem~\ref{oscexist} uses the fact that, if $\tau_c=\infty$, then the solution must be oscillating.  The converse is also true.
\begin{Crly}\label{osctauinf}
If $(A,\rw)$ is an oscillating solution, then $\tau_c=\infty$, $\Phi_c=0$, and $2\Lambda r_c^2\le 1$.
\end{Crly}
\noindent
\textbf{Proof}:  We shall prove shortly that $\kappa_c=-\infty$ whenever $\lf_{\tau\nearrow \tau_c}\kappa(\tau)=-\infty$.  In this case, it follows readily from Equation~(\ref{Theq}) that the solution cannot be oscillating.  If $\kappa_c>-\infty$, since the equations~(\ref{req})~-~(\ref{keq}) are nonsingular at $\tau_c$ whenever $\tau_c<\infty$, $\rw$ can equal zero on a sequence approaching $\tau_c$ only if $\rw\equiv 0$ in a neighborhood of $\tau_c$.  Differentiating Equations~(\ref{req}),~(\ref{Neq}), and~(\ref{keq}) indefinitely gives $r$ constant which is impossible.  Thus, for an oscillating solution, $\tau_c$ must be infinite.  It follows from the proof of Theorem~\ref{smbag} that $\Phi_c=0$.  Finally, Equation~(\ref{keq}) yields $2\Lambda r_c^2\le 1$ whenever $\kappa_c$ is finite and $\tau_c=\infty$.  \hfill $\blacksquare$\\ \\
(2) For oscillating solutions, Equations~(\ref{Ctau}) and~(\ref{taumet}) imply that the spacetime has infinite volume between $r=0$ and $r=r_c$.  Consequently, the solution between $\tau=-\infty$ and $\tau=\infty$ should be viewed as complete.\\ \\
(3)  Oscillating solutions cannot occur for large $\Lambda$.  In fact, we have the following:\\[10cm]
\begin{thm}\label{osclsmall}
For any oscillating solution, $\Lambda\le 1/4$ and $r_c\le \sqrt{2}$.
\end{thm}
\noindent
\textbf{Proof}:  For any oscillating solution, $2\Lambda r_c^2\le 1$, $\Phi_c=0$, and $\rw^2_c=0$ or $1$.   Clearly, $\Phi_c=0$ while $\rw^2_c=1$ is impossible.  The following quadratic in $r_c^2$
\begin{equation}\label{pquad}
0=r_c^2\Phi_c=r_c^2-1-\Lambda r_c^4
\end{equation}
gives $\Lambda\le 1/4$.  Equation~(\ref{pquad}) and $2\Lambda r_c^2\le 1$ yield $r_c\le \sqrt{2}$.\hfill $\blacksquare$

\section{$2\Lambda r_c^2\ge 1$}
\begin{thm}\label{wbiggen}
For any value of $2\Lambda r_c^2$,  the solution is generic whenever $\rw^2_c>1$.
\end{thm}
\noindent
\textbf{Proof}:  Without loss of generality, we assume $\rw_c>1$.  Our first goal is to prove that Equations~(\ref{req})~-~(\ref{keq}) are nonsingular at $\tau_c$, which is finite.  We have already established that $\rw_c<\infty$ whereas $\rw'_c=\infty$.   We now claim that $\Phi_c<0$.  If, to the contrary, $\Phi_c=0$, then differentiating Equation~(\ref{Phi}) gives $\Phi>0$ for $r$ near to but less than $r_c$.  Thus, $\lf_{\tau\nearrow \tau_c}\kappa(\tau)\ge 0$. It is clear from Equation~(\ref{feq}) that $f$ is bounded near $r_c$ and from Equation~(\ref{k}) that $\lp_{\tau\nearrow \tau_c}\kappa(\tau)<\infty$.  Equation~(\ref{k}) then gives $U_c=0$.  However, Equation~(\ref{Ueq}) implies $U_c$ cannot be zero.  Thus $\Phi_c<0$.  Corollary~\ref{tcfin} now gives $\tau_c<\infty$.

Next, we prove that $U_c>0$ and use this to prove that $\kappa$ is bounded.  $2(1-A-2(1-\rw^2)^2/r^2)$, the rightmost term on the left side of Equation~(\ref{zeq}) has a limit as $r\nearrow r_c$.  Either this limit is nonzero or the term approaches zero from above.  In either case, there is a neighborhood of $r_c$ where this term is always positive or always negative.  It follows from Equation~(\ref{zeq}), as in Lemma~\ref{limits}, that $z$ has a finite limit $z_c$.  If $z_c<0$, then Equation~(\ref{aw'beta}), with small positive $\epsilon$, gives $\lim_{r\nearrow r_c}(A{\rw'}^{2-\epsilon})'(r)=+\infty$ which is impossible since the finiteness of $f_c$ gives $A{\rw'}^{2-\epsilon}\rightarrow 0$.  Thus $z_c\ge 0$.  Since $\Phi_c<0$, Equation~(\ref{z}) gives $+\infty>U_c=+\sqrt{f_c}>0$ as desired.

To prove $\kappa$ is bounded, we first notice that it is obvious from Equation~(\ref{keq}) that $\lp_{\tau\nearrow\tau_c}\kappa(\tau)<\infty.$  If we assume $\lf_{\tau\nearrow \tau_c}\kappa(\tau)=-\infty$, then Equation~(\ref{keq}) also gives
\begin{equation}\label{kliminf}
\lim_{\tau\nearrow \tau_c}\kappa(\tau)=-\infty
\end{equation}
and
\begin{equation}\label{krec}
\lim_{\tau\nearrow \tau_c}\frac{\rd}{\rd \tau}(\frac{1}{\kappa})=1.
\end{equation}
We differentiate once more with respect to $\tau$ to obtain
\begin{equation}\label{1/kddot}
\lim_{\tau\nearrow \tau_c}\frac{\rd^2}{\rd \tau^2}(\frac{1}{\kappa})=0.
\end{equation}
Consequently, for every $\epsilon>0$, there exist $\delta$ such that whenever $0<\tau_c-\tau<\delta$,
\begin{equation}\label{kbnd}
\kappa(\tau)<\frac{1}{\tau-\tau_c}+\frac{\epsilon}{1-\epsilon(\tau-\tau_c)}.
\end{equation}
Integrating Inequality~(\ref{kbnd}) gives
\begin{equation}\label{kint}
\int_{\tau_c-\delta}^{\tau_c}\kappa(\tilde \tau)\rd \tilde\tau=-\infty.
\end{equation}
Substituting Inequality~(\ref{kint}) into Equation~(\ref{Ueq}) gives $U_c=+\infty$ which is impossible.  Therefore $\kappa$ is bounded.

Since $\kappa$, $U$, and $\rw$ are all  bounded, Equations~(\ref{req})~-~(\ref{keq}) are nonsingular at $\tau_c$. We extend the solution beyond $\tau_c$ and consider the corresponding solution of Equations~(\ref{Aeq}) and~(\ref{weq}).  Because $N$ changes sign at $\tau_c$ the corresponding solution of Equations~(\ref{Aeq})~ and~(\ref{weq}) is in a region of $r$ decreasing from $r_c$ to some $r_0$ where either $r_0=0$ or the Equations become singular at $r_0$.  Also, since $U$ does not change sign at $\tau_c$, $\rw'(r)<0$ near $r_c$ for the extended solution. 

From Equation~(\ref{weq}) it is obvious that $\rw'<0$ for all $r>r_0$.  Finally, we claim that the solution cannot be a black hole.  Indeed, otherwise there exists a sequence $\{r_n\}\searrow r_0>0$ such that $\lim_{n\nearrow \infty}A(r)=0$.  This sequence can be chosen so that $A'$ and hence $\Phi$ are positive on it.  Since $\Phi$ has a limit $\Phi_0$ as $r\searrow r_0$,  $\Phi_0=\lim_{r\searrow r_0}\Phi(r)\ge 0$; i.e., $\rw_0=\lim_{r\searrow r_0}\rw(r)$ is finite and $r_0$ cannot equal $0$.  
Furthermore, it is clear from Equation~(\ref{veq}) that $A\rw'$ has a nonzero limit as $r\searrow r_0$.  Thus, $f_0=\lim_{r\searrow r_0}f(r)=\infty$.  Equation~(\ref{Aeq}) gives $\lim_{r\searrow r_0}A'(r)$, preventing $A$ from going to zero. 

Because $\rw_0>1$, the solution cannot be smooth at $r=0$ either.  The only possibility remaining is that the original solution is generic.\hfill $\blacksquare$\\[10cm]
\begin{Lma}\label{kinf}
Unless the solution is noncompact, $\kappa_c>-\infty$.
\end{Lma}
\noindent
\textbf{Proof}:  We may assume $\lp_{r\nearrow r_c}\rw^2(r)\le 1$.  We also assume $\lf_{\tau\nearrow \tau_c}\kappa(\tau)=-\infty$.  As in the proof of Theorem~\ref{wbiggen}, because $\Phi$ is bounded, Equation~(\ref{feq}) gives $U^2=f$ is also bounded.  We compare Equation~(\ref{keq}) to the equation
\begin{equation}\label{xeq}
\dot x=-\frac{x^2}{2}
\end{equation}
which has unique solution
\begin{equation}\label{xsol}
x(\tau)=\frac{2\bar x}{2+\bar x(\tau-\bar\tau)}
\end{equation}
through the point $(\bar\tau,\bar x)$.  It is clear that we can choose $\bar\tau$ and $\bar x=\kappa(\bar\tau)<0$ such that for all $\tau\in(\bar\tau,\tau_c)$, $\dot\kappa<\dot x$ and, consequently, $\kappa<x$.  Therefore, $\tau_c\le \bar\tau-2/\bar x<\infty$.
  
Equation~(\ref{keq}) also gives $\lim_{\tau\nearrow \tau_c}\kappa(\tau)=-\infty$ and Inequality~(\ref{kbnd}),
\[\kappa(\tau)<\frac{1}{\tau-\tau_c}+\frac{\epsilon}{1-\epsilon(\tau-\tau_c)}.\]
Substituting Inqualities~(\ref{kbnd}) into Equation~(\ref{Neq}) yields
\begin{equation}\label{Neqest}
\dot N<[\frac{1}{\tau-\tau_c}+\frac{\epsilon}{1-\epsilon(\tau-\tau_c)}]N.
\end{equation}
Choosing any $\bar\tau$ sufficiently close to $\tau_c$ such that Inequality~(\ref{kbnd}) holds and integrating Inequality~(\ref{Neqest}) from $\bar\tau$ to $\tau_c$ gives constants $c>0$ such that, in this region,
\begin{equation}\label{Nest}
N(\tau)>c(\tau_c-\tau).
\end{equation}
Inequalities~(\ref{kbnd}) and~(\ref{Nest}) give
\begin{equation}\label{kNneg}
\lp_{\tau\nearrow \tau_c}2\kappa N(\tau)=\lp_{r\nearrow r_c}rz(r)<0.
\end{equation}
Furthermore, substituting Inequality~(\ref{kbnd}) into Equation~(\ref{Ueq}) gives $U_c=0$ since otherwise $U_c$ is unbounded. From Equation~(\ref{wtaueq}) it follows that $\Phi$ has a limit $\Phi_c$.  Equation~(\ref{kNneg}) implies that this limit must be negative.

We have $\Phi_c<0$ and $U_c=0$.  To finish the proof that the solution is noncompact, it suffices to prove that $\lp_{r\nearrow r_c}{\rw'}^2(r)<\infty$ \cite{L2}.  From Equation~(\ref{weq}), it is clear that if $\rw$ becomes infinite on a sequence, then $\lim_{r\nearrow r_c}{\rw'}^2=\infty$.  However, Equation~(\ref{feq}) shows that ${\rw'}^2$ cannot become infinite while $f_c=0$.\hfill $\blacksquare$\\

The local nature of solutions near the horizon can be summarized as follows:
\begin{thm}\label{nonsing}
For all solutions of Equations~(\ref{req})~-~(\ref{keq}) except for noncompact and oscillating solutions, there exists a finite $\tau_c$, where the equations are nonsingular and $N$ changes sign.
\end{thm}
Thus, if a solution is neither noncompact nor oscillating, $r$ decreases in the extended solution beyond the singularity at $r_c$.  The only possibilities are that $A>0$ for all $r<r_c$ in this extended region, in which case the solution is either generic or compact, or there exists some $r_0\in(0,r_c)$ such that $A(r_0)=0$.  In this latter case, we  will prove that the solution is a black hole.  In other words, every solution is one of our five types.   We have already established the existence of solutions of each type except for black holes.  What remains is to prove the existence of black holes solutions.  The first step towards this is a refinement of Corollary~\ref{rc<1}, which is interesting in its own right:
\begin{thm}\label{rcrsmall}
For any $\tilde r>0$, there exist $\tilde\lambda\ge 2$ such that for all $\Lambda\ge 0$ and $\lambda>\tilde\lambda$,  $r_c<\tilde r$.
\end{thm}
\noindent
\textbf{Proof}:  For any $c>2$, as in Lemmas~\ref{hneg} and~\ref{g'neg}, we define
\begin{equation}\label{gc}
g_c(r)=cr^2-(1-\rw^2).
\end{equation}
Simple calculations yield, for any solution of Equations~(\ref{Aeq}) and~(\ref{weq})
\begin{equation}\label{gceq}
g_c''(r)=-\frac{\Phi}{rA}g_c'+\frac{2}{r^2A}[r^2A{\rw'}^2+(1-\rw^2)(1-c+\rw^2)+g-c\Lambda r^4].
\end{equation}
$g_c(0)=g_c(0)=0$ and $g_c''(0)=2(c-\lambda)<0$ whenever $\lambda>c$.

Now clearly, whenever $g_c<0$, $g_2=g<0$ also.  Consequently, as in Lemma~\ref{hneg}, for any $\Lambda\ge 0$,  $h(r)<0$ whenever $g_c<0$.  Also, as in Lemma~\ref{g'neg}, if $\tilde r$ is the smallest $r$ such that $g_c(r)=0$, then 
\begin{eqnarray}\label{gc''comp}
g_c''(\tilde r)&=&\frac{2}{r^2A}[r^2A{\rw'}^2-(1-\rw^2)^2+g-c(1-\rw^2)-c\Lambda r^4]_{r=\tilde r}\nonumber\\
&\le&\frac{2}{r^2A}[r^2\rw^2{\rw'}^2-(1-\rw^2)^2+g]_{r=\tilde r}\nonumber\\
&=&\frac{2}{r^2A}[c^2 r^4-(1-\rw^2)^2+g]_{r=\tilde r}\nonumber\\
&=&\frac{2}{r^2A}[g_c(1+cr^2+(1-\rw^2))]_{r=\tilde r}<0
\end{eqnarray}
whenever $h(\tilde r)$ and $g_c(\tilde r)$ are negative.  Consequently, we have
for any $\Lambda\ge 0$, $g_c'(r)<0$ as long as $h(r)<0$.  

If, $r_c\ge 1/\sqrt{c}$, then $g_c(r_c)>0$.  It follows that there exists a smallest $r_2$ such that $g_c'(r_2)=0$.  Therefore, $h(r)<0$ for all $r<r_2+\delta_1$ for some small positive $\delta_1$ and $g_c'(r)<0$ for all $r<r_2+\delta_2$ for some small positive $\delta_2$.  This contradicts the fact that $g_c'(r_2)=0$. \hfill $\blacksquare$
\begin{thm}\label{lbigno0}
For any $\Lambda>0$, whenever $\lambda\ge\max\{2\Lambda,2\}$, the solution is generic and $\rw$ has no zeros.
\end{thm}
\noindent
\textbf{Proof}:  If $\Lambda\le 1$ and $\lambda>2$, then $2\Lambda r_c^2<1$.  If $\Lambda>1$ and $\lambda\ge 2\Lambda$, then $2\Lambda r_c^2<1$ also.  Thus, $\kappa>0$ always.  It follows from Corollary~\ref{bog} that the solution is generic.

We have already proved that $\rw$ cannot have any zeros in the interval $\tau\in(-\infty,\tau_c)$.  We now suppose that $\rw$ has a zero at $\tilde\tau>\tau_c$and will arrive at a contradiction.

We choose $\tau_1>\tau_c$ such that $-(N+\rw)(\tau_1)<0$ and $\rw(\tau_1)>0$.  We also choose $\tau_2\in(\tilde\tau,\tau_0)$ such that $\rw(\tau_2)<0$.  Next, we define
\begin{equation}\label{Nmin}
\bar{N}=\max_{\tau\in[\tau_1,\tau_2]} \{N(\tau)\},
\end{equation}
\begin{equation}\label{kmax}
\bar\kappa=\max_{\tau\in[\tau_c,\tau_2]}\{\kappa(\tau)\},
\end{equation}
and
\begin{equation}\label{u2max}
\bar{u}^2=\max_{\tau\in[\tau_c,\tau_2]}\{u^2(\tau)\}.
\end{equation} 
It follows immediately from Equation~(\ref{Neq}) that $\dot{N}<0$ for all $\tau>\tau_c$ and thus, obviously, whenever $\tau>\tau_1$, $N(\tau)<N(\tau_1)$.  Next, for any $\alpha>0$, we define
\begin{equation}\label{halpha}
h_\alpha(r)=-r^\alpha N-\rw.
\end{equation}
A simple calculation yields, for every solution of Equations~(\ref{req})~-~(\ref{keq}),
\begin{equation}\label{halpha'}
\dot{h_\alpha}=r^\alpha[N^2(1-\alpha)-\kappa N+2u^2].
\end{equation}
Clearly, since $r<1$, for any $\alpha>1$, $h_{\alpha}(\tau_1)<0$.  Also, $h_{\alpha}(\tilde\tau)>0$.  Consequently, there exists a $\bar\tau_\alpha\in(\tau_1,\tilde\tau)\subset(\tau_1,\tau_2)$ such that $h_\alpha(\bar\tau_\alpha)>0$.  However, whenever
\[\alpha>1+\frac{2\bar u^2-\bar\kappa N(\tau_2)}{\bar N^2},\]
$\dot{h_\alpha}<0$ for all $\tau\in(\tau_1,\tau_2)$.  The result follows.\hfill $\blacksquare$ 
\begin{thm}\label{Lbigno0}
Whenever $\Lambda\ge3$, $\rw$ has no zeros.
\end{thm}
\noindent
\textbf{Proof}: We recall Equation~(\ref{a-w2}) from Lemma~\ref{hneg}
\[h(r)=A-\rw^2.\]
Since $2\lambda(1-\lambda)\le 1/4$, whenever $\Lambda>3/4$, $h''(0)<0$.  Consequently, $h<0$ for small $r$.  As in Lemma~\ref{hneg}, we have, for any $\tilde r$ such that $h(\tilde r)=0$, 
\begin{equation}\label{h'h0}
h'(\tilde r)=\frac{\Phi}{r}-\frac{2s^2}{r}-2s
\end{equation}
where $s=\rw\rw'$.  Considering the right side of Equation~(\ref{h'h0}) as a quadratic in $s$ gives discriminant
\begin{equation}\label{Delta}
\Delta=4[1+\frac{2(1-\rw^2)}{r^2}(1-\frac{1-\rw^2}{r^2})-2\Lambda].
\end{equation}
Now, 
\[\frac{2(1-\rw^2)}{r^2}(1-\frac{1-\rw^2}{r^2})\le 1/2.\]
Consequently, if $\Lambda>3/4$, $\Delta<0$.  As in Lemma~\ref{a-w2}, there can be no $\tilde r$ such that $h'(\tilde r)\ge 0$ while $h(\tilde r)=0$.  It follows that $\rw>0$ for all $r\le r_c$.  Since $r_c\le \sqrt{3/\Lambda}$ with equality only for the Reissner-Nordstr\"om solution, arguments similar to those used in Theorem~\ref{lbigno0} prove that, whenever $\Lambda\ge 3$, $\rw$ has no zeros in the interval $(\tau_c,\tau_0)$ either.\hfill $\blacksquare$
\begin{thm}\label{bh}
For any positive integer $n$, there exist black hole solutions in which $\rw$ has at least $n$ zeros.
\end{thm}
We consider $L=[0,\Lambda_0]\times [0,\max\{2,2\Lambda_0\}]$ with the standard topology and all solutions subject to the condition that $(\Lambda,\lambda)\in L$.  We recall that there exists an interval $l_n=(\lambda_n,\lambda_{n+1})$ such that whenever $\Lambda=0$ and $\lambda\in l_n$, the solution has exactly $n+1$ zeros \cite{jS93}.  $\lambda_n$ produces the $n^{\mathrm{th}}$ particlelike solution and $\lambda_{n+1}$ produces the $(n+1)^{\mathrm{th}}$.  

We denote by $W_n$ the subset of $L$ such that w of the solution with $(\Lambda,\lambda)\in L$ has exactly $n$ zeros and by $C_n$ the closure of the path connected component of $W_n$ that contains $l_n$ minus the set $\{0\}\times l_n$.  Because $C_n$ is compact, its boundary $\partial C_n$ must be connected.  From Theorems~\ref{lbigno0} and~\ref{Lbigno0} and results from \cite{jS93} it follows that 
\begin{equation}\label{bndint}
\partial C_n\cap \partial L=\{(0,\lambda_n),(0,\lambda_{n+1})\}.
\end{equation}
Next, we define $\bar\partial C_n$ as the set of points in $L$ that satisfy the following: $(\Lambda,\lambda)\in \bar\partial C_n$ whenever

(i)  $(\Lambda,\lambda)\in \partial C_n$,

(ii)  the solution with this choice of parameters is noncompact, and

(iii) w in this solution has $n$ zeros.\\ \\

Similarly, we define $\bar\partial C_{n+1}$ as the set of points in $L$ that satisfy the following: $(\Lambda,\lambda)\in \bar\partial C_{n+1}$ whenever

(i)  $(\Lambda,\lambda)\in \partial C_n$,

(ii)  the solution with this choice of parameters is noncompact, and

(iii) w in this solution has $n+1$ zeros.\\ \\
(See Figure~1.) $\bar\partial C_{n+1}$ is nonempty since it contains $(0,\lambda_{n+1})$.  We claim that for any $(\Lambda,\lambda)\in\bar\partial C_{n+1}$, there exists a neighborhood $V$ such that for each point in $V$, the corresponding solution has at least $n+1$ zeros unless there exists a black hole solution in $V$.  Because $\bar\partial C_n$ is also nonempty (it contains $(0,\lambda_n)$) and $\partial C_n$ is connected, the result follows. \hfill $\blacksquare$
\begin{thm}\label{nearnc}
Let $(\hat\Lambda,\hat\lambda)\in\bar\partial C_{n+1}$.  There exists a neighborhood $V$ of $(\hat\Lambda,\hat\lambda)$ such that for each $(\Lambda,\lambda)\in V$, one of the following holds:\\

(A) the solution is generic and $\rw$ has at least $n+1$ zeros,

(B) the solution is noncompact and $\rw$ has exactly $n+1$ zeros, or

(C) the solution is a black hole.\\ \\
In particular, $V\cap\bar\partial C_n$ is empty.
\end{thm}
Noting that for noncompact solutions, $\kappa_c=-\infty$, we prove Theorem~\ref{nearnc}, by using the following lemmas:
\begin{Lma}\label{kstaysbig}
Let $\kappa_0<-(3+\sqrt{5})/2$ and suppose there exist $\tilde\tau \in(-\infty,\tau_0)$ such that $\kappa(\tilde\tau)<\kappa_0$.  Then either there exist $\tau_1\in(\tilde\tau,\tau_0)$ such that $N^2(\tau_1)>1$ or $\kappa(\tau)<\kappa_0+1$ for all $\tau\in(\tilde\tau,\tau_0).$
\end{Lma}
\noindent
\textbf{Proof}:  We assume $-1\le N\le 1$ for all $\tau\in (-\infty,\tau_0)$ and consider the following equation which follows easily from Equations~(\ref{Neq}) and~(\ref{keq}):
\begin{equation}\label{k+N}
\dot\kappa+\dot N=1-\kappa^2-2\Lambda r^2+(\kappa-N)N.
\end{equation}
Clearly, $(\dot\kappa+\dot N)<1-\kappa^2-\kappa$ which is negative whenever $\kappa<-(1+\sqrt{5})/2$.  The result follows.\hfill $\blacksquare$
\begin{Lma}\label{rckcont}
At any $(\hat\Lambda,\hat\lambda)$ that yields a noncompact solution, $r_c$ is a continuous function of $\Lambda$ and $\lambda$.  Also, $\kappa_c$ is a continuous function of $\Lambda$ and $\lambda$ in the extended reals.
\end{Lma}
\noindent\textbf{Proof}: For noncompact solutions, $\kappa_c=-\infty$.  Because of continuous dependence on parameters, to establish the continuity of $r_c$ and $\kappa_c$ it suffices to establish, for small positive $\epsilon$, the existence of positive $\delta$ and $M$ such that $(\sqrt{A})'(\lambda,r)<-M$ whenever $0\le \max\{|\hat\lambda-\lambda|,|\hat\Lambda-\Lambda|\}<\delta$ and $r\in[r_c(\hat\Lambda,\hat\lambda)-\epsilon,r_c(\hat\Lambda,\hat\lambda))$.  Now, Equations~(\ref{req}) and~(\ref{Neq}) give 
\begin{equation}\label{sAeq}
(\sqrt{A})'=\frac{\kappa-N}{r}-\frac{2U^2}{rN}<\frac{\kappa-N}{r}.
\end{equation}  
It follow easily that $(\sqrt{A})'_c=-\infty$.  Also, for arbitrarily large $M$ and sufficiently small $\epsilon>0$, there exist positive $\delta$ such that $(\sqrt{A})'(\Lambda,\lambda,r_c(\Lambda,\lambda)-\epsilon)<-M$ whenever $0\le\max\{|\Lambda-\tilde\Lambda|,|\lambda-\tilde\lambda|\}<\delta$.  Since for any $\Lambda$ and $\lambda$ and all $\tau\in(0,\tau_c(\Lambda,\lambda))$, $N\le 1$, Lemma~\ref{kstaysbig} gives
\begin{equation}\label{kbig}
\kappa(\Lambda,\lambda,\tau)<-M+1
\end{equation}
for all $(\Lambda,\lambda)$ such that $\max\{|\Lambda-\hat\Lambda|,|\lambda-\hat\lambda|\}<\delta$ and all $\tau\in(\tau(r_c(\hat\Lambda,\hat\lambda))-\epsilon,\tau_c(\Lambda,\lambda))$. \hfill $\blacksquare$
$\;$\\ \\ \\ \\ \\ \\ \\ \\ \\ \\ \\ \\ \\ \\ \\ \\
\begin{flushright}
\begin{figure}[b]\label{map}
\setlength{\unitlength}{3pt}
\begin{picture}(100,20)(0,0)
\put (0,0){\vector(1,0){105}}
\put (0,0){\vector(0,1){105}}
\put (0,100){\line(1,0){100}}
\put (100,0){\line(0,1){100}}
\put (107,-0.5){$\Lambda$}
\put (-3,85){$\lambda$}
\put (33,-0.5){\line(0,1){1}}
\put (30,-3){$1/4$}
\put (99,-3){$3$}
\put (-3,99.5){$2$}
\qbezier(0,75),(10,76),(33,72)
\put (10,77){oscillating solutions}
\qbezier(0,45)(25,42),(46.7,46.4)
\qbezier(0,65),(25,68),(50,63)
\put(57,60){\circle*{0.8}}
\put(46.7,46.4){\circle*{0.8}}
\put(46.7,46.4){\circle{10}}
\put(57,60){\circle{10}}
\put(57.3,61.5){V}
\put(20,66){$\bar\partial C_{n+1}$}
\put(60,52.5){Black holes}
\put(18.5,44){$\bar\partial C_n$}
\put(20,54){$C_n$}
\qbezier(0,20)(20,15)(56,49)
\put(1,53){$l_n$}
\put(-4,45){$\lambda_n$}
\put(-8,65){$\lambda_{n+1}$}
\put(14,32){$C_{n-1}$}
\put(67,25){\circle*{0.8}}
\put(69,24){Einstein Space}
\put(50,55){\oval(16,16)[r]}
\put(46.7,46.4){\line(6,1){3.4}}
\end{picture}
\caption{The geometry of solutions.  Black hole solutions are on the open set contained between the centers of the two circles. $\partial C_{n+1}$ is the portion of the boundary of $C_n$ between $\Lambda=0$ and the center of V.  There must be a $V$ that contains a black hole solution.}
\end{figure}
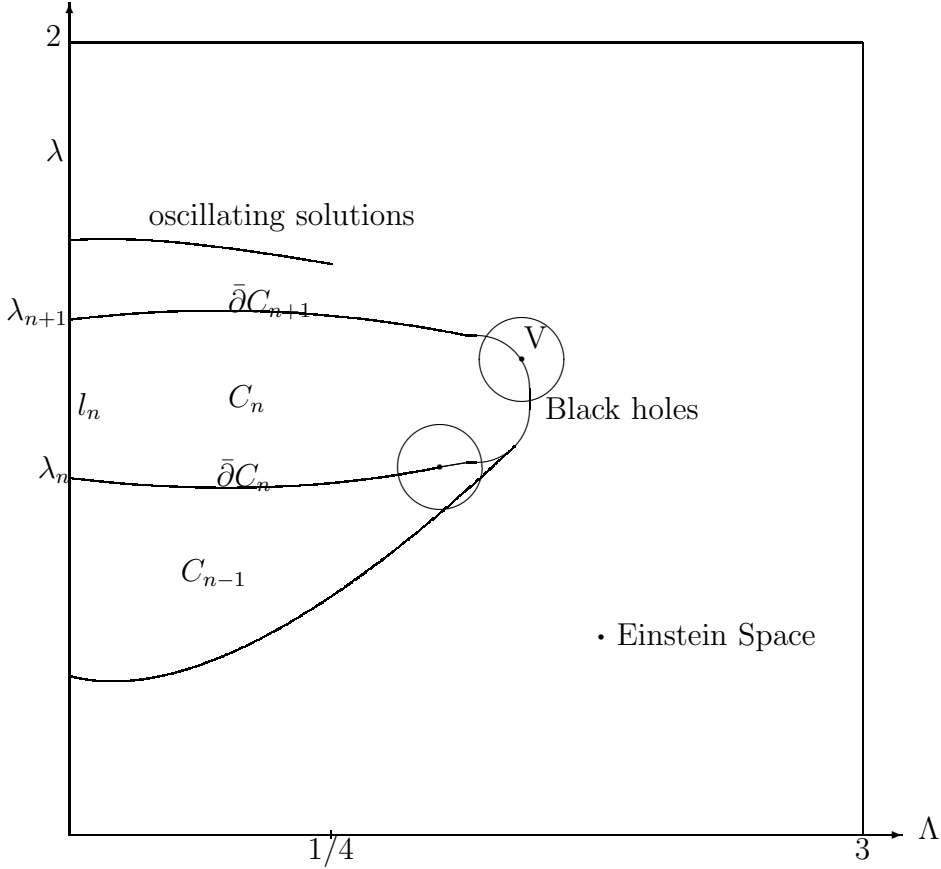
\end{flushright}
\begin{Lma}\label{ncwn0}
If $(A,\rw)$ is a noncompact solution, then $\rw_c\ne 0$.
\end{Lma}
\noindent
\textbf{Proof}:  Equation~(\ref{weq}) implies that 
\[\rw'(\hat r)=\frac{\rw(1-\rw^2)(\hat r)}{
\hat r\Phi(\hat r)}\]
at any $\hat r$ such that $\rw''(\hat r)=0$.  Since $\Phi_c<0$ for noncompact solutions, taking any sequence of such $\hat r$ approaching $r_c$ establishes that $\rw'$ has a limit $\rw'_c$ as $r\rightarrow r_c$.  We assume $\rw_c=0$ and consider the following two possibilities:\\

(1) $\rw'_c\ne 0$ and

(2) $\rw'(c)=0$.\\ \\
\textit{Case} 1.  $\Phi_c<0$ whereas $f_c=0$.  This contradicts Equation~(\ref{feq}).\\
\textit{Case} 2.  Equation~(\ref{weq}) gives
\begin{equation}\label{w'byw}
(\frac{\rw'}{\rw})'+(\frac{\rw'}{\rw})^2+\frac{\Phi}{rA}(\frac{\rw'}{\rw})+\frac{1-\rw^2}{r^2A}=0.
\end{equation}
Setting $(\rw'/\rw)'=0$ and considering the remaining equation as a quadratic in $(\rw'/\rw)$ gives
\begin{equation}\label{wbyw'0}
\frac{\rw'}{\rw}=-\frac{\Phi}{2rA}\pm\frac{1}{2rA}\sqrt{\Phi^2-4A}.
\end{equation}
Expanding in $A$ about $A=0$ yields
\begin{equation}\label{w'bywexp} 
\frac{\rw'}{\rw}=-\frac{\Phi}{2rA}\pm(\frac{\Phi}{rA}-\frac{1}{r} +\circ\;A^2).
\end{equation}
Consequently, on any sequence $r_n\nearrow r_c$ such that $(\rw'/\rw)'=0$, either $(\rw'/\rw)(r_n)\rightarrow \infty$ or $(\rw'/\rw)(r_n)\rightarrow -1/r_c$.  Now Equation~(\ref{Theq}) implies that the solution cannot be oscillating; i.e., $\rw'$ has only one sign near $r_c$.  It follows that $\rw'/\rw$ has a limit as $r\nearrow r_c$.  Since $\ln|w|\rightarrow 0$ as $r\nearrow r_c$, $\rw'/\rw$ must diverge.  Finally, dividing Equation~(\ref{feq}) by ${\rw'}^2$ yields $f'>0$ near $r_c$.  However, since $f>0$ and $f_c=0$, this is impossible.\hfill $\blacksquare$\\[10cm]
\textbf{Proof of Theorem~\ref{nearnc}}:  There exists a neighborhood $V$ of $(\hat\Lambda,\hat\lambda)$ such that $r_c$ varies continuously with $\Lambda$ and $\lambda$.  It follows from Equation~(\ref{Theq}) that because $\kappa_c\ll 0$ in $V$, $V$ can contain no oscillating solutions.  Also, Lemma~\ref{ncwn0} excludes the possibility that $V$ contains any noncompact solutions in which $\rw$ has any other than $n+1$ zeros.  All other solutions behave locally like generic solutions near $r_c$.  Thus, by continuous dependence on parameters and the uniqueness of solutions satisfying $\rw(\tilde r)=\rw'(\tilde r)=0$, $A(\tilde r)>0$ for any $\tilde r$, it follows that for no solution in $V$ can $\rw$ have fewer than $n+1$ zeros. \hfill $\blacksquare$\\

The figures below show the geometrical changes in solutions near a black hole solution.  Each figure below is a plot of $\rw$ versus $r$.  Figure~2 is a black hole solution.  The following figures are perturbations of this solution and exhibit different geometries.
\pagebreak
$\;$\\
\begin{figure}[t]\label{BH2}
\epsfig{figure=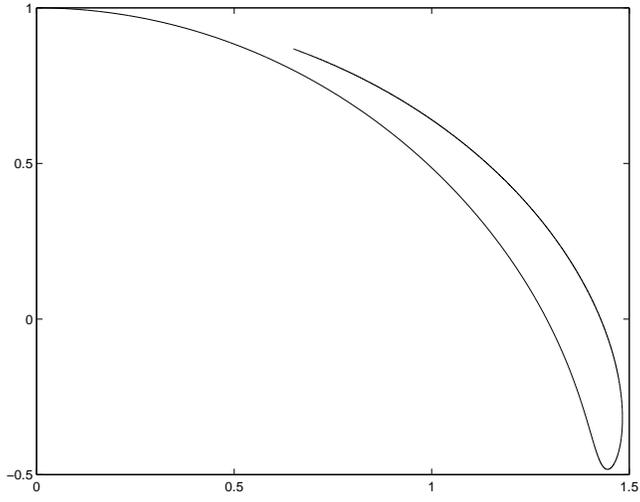,scale=.5}
\caption{Black hole solution with 2 zeros.  $\Lambda=0.3394$, $\lambda=0.9075$.}
\end{figure}
\bigskip
\pagebreak
$\;$\\
\begin{figure}[t]\label{gen2}
\epsfig{figure=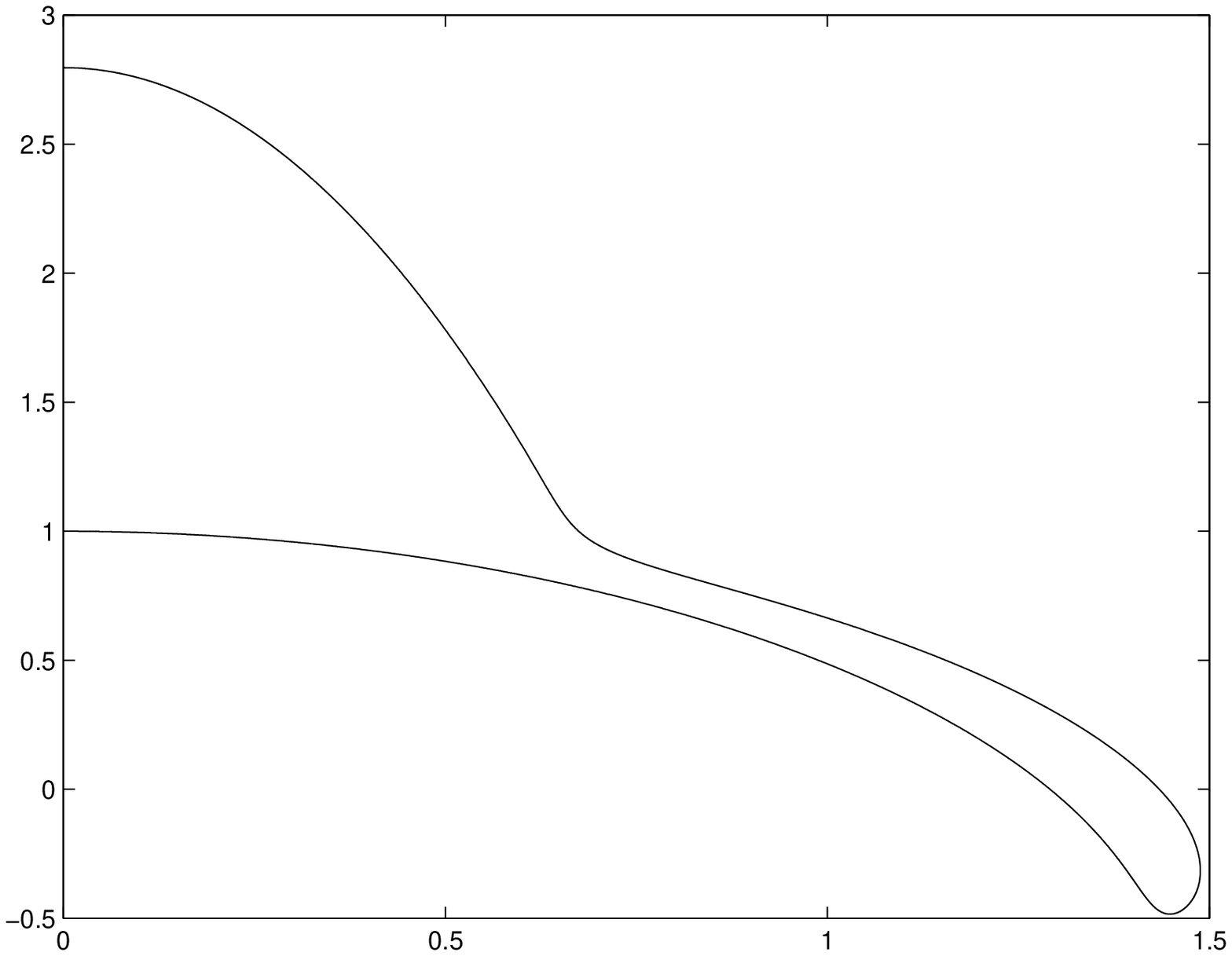,scale=.5}
\caption{Generic solution with 2 zeros.}
\end{figure}
\smallskip
\pagebreak
$\;$\\
\begin{figure}[t]\label{gen3}
\epsfig{figure=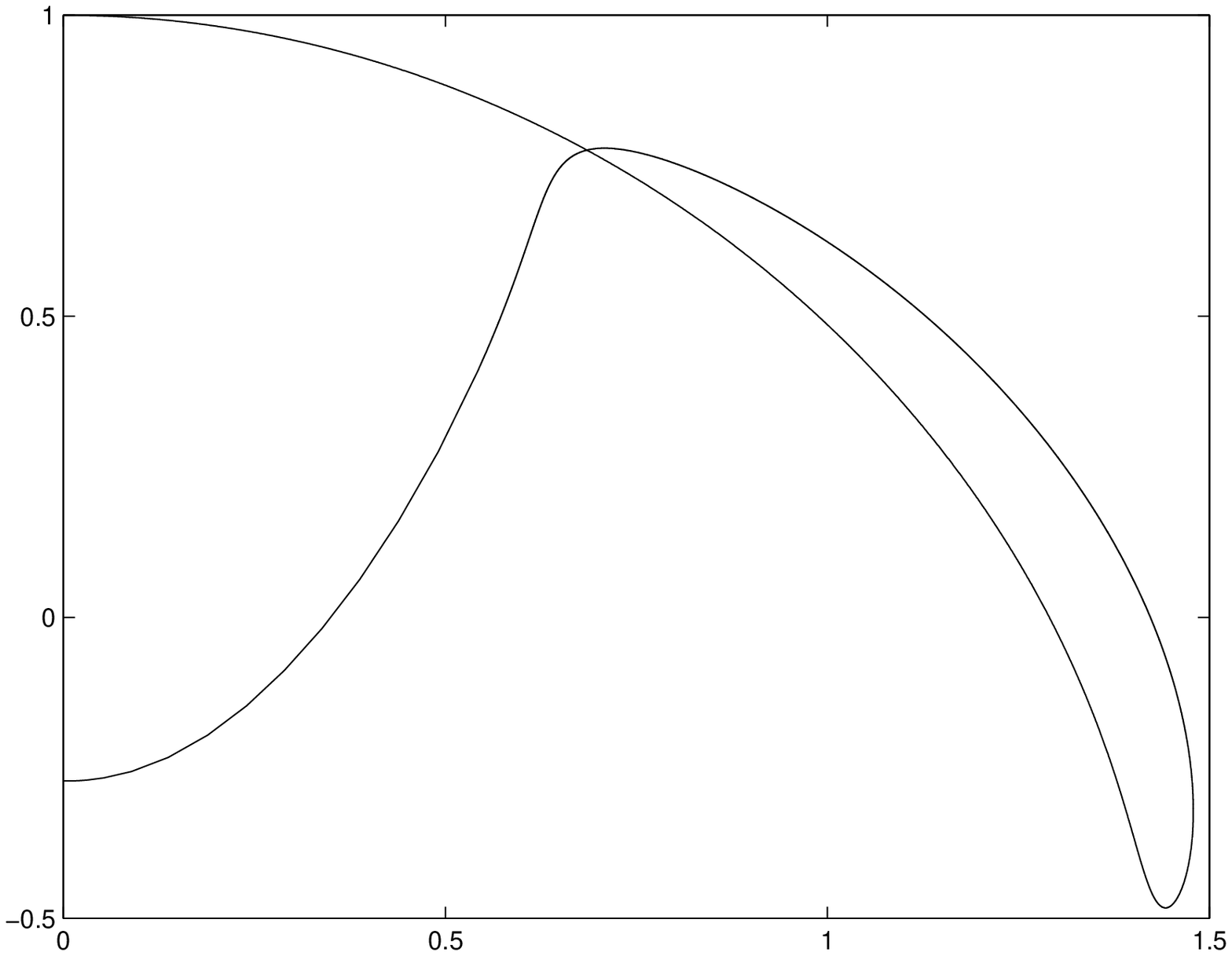,scale=.5}
\caption{Generic solution with 3 zeros}
\end{figure}
\pagebreak
$\;$\\
\begin{figure}[t]\label{nc1}
\epsfig{figure=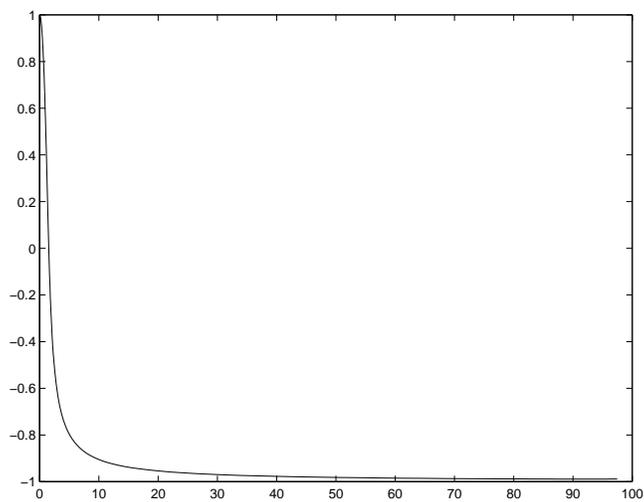,scale=.5}
\caption{Noncompact solution with 1 zero.  $\Lambda=3\times 10^{-4}$, $\lambda=0.9074$.}
\end{figure}
\bigskip
\pagebreak
$\;$\\
\begin{figure}[t]\label{nc2}
\epsfig{figure=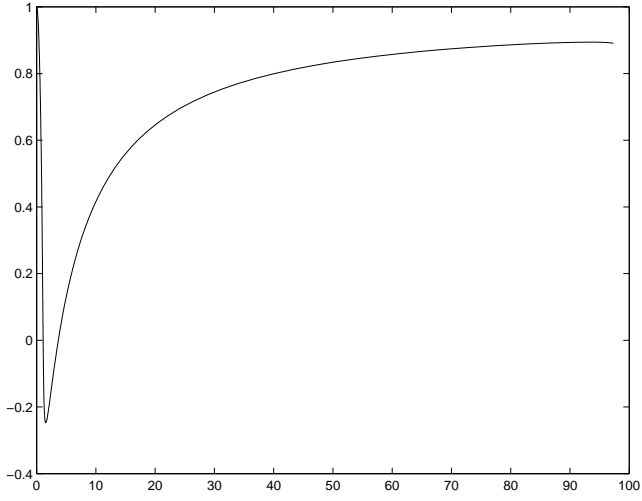,scale=.5}
\caption{Noncompact solution with 2 zeros.  $\Lambda=3\times 10^{-4}$, $\lambda=1.3047$.}
\end{figure}
\bigskip

We conclude by describing the behavior of solutions near a black hole singularity at $r_0$.  Recall that $r_0$ is the value of $r$ at which the extended solution crashes with $A=0$.
\begin{Lma}\label{bhkinf}
Suppose there exists a sequence $\{\tau_n\}\nearrow \tau_0$ such that $N(\tau_n)\nearrow 0$ and $r_0=r(\tau_0)>0$.  Then $\Phi$ and $f$ have limits $\Phi_0$ and $f_0$ respectively and $f_0=0$.
\end{Lma}
\noindent
\textbf{Proof}:  We first prove that $U$ is bounded as $\tau\nearrow \tau_0$; or, equivalently, that $A{\rw'}^2$ is bounded as $r\searrow r_0$.  Indeed, we  
recall Equation~(\ref{feq})
\[r^2(A{\rw'}^2)'+r{\rw'}^2[\Phi+2A{\rw'}^2]+2\rw\rw'(1-\rw^2)=0.\]
For any $\epsilon>0$ and any sequence $\{r_n\}\searrow r_0$ such that $A{\rw'}^2>\epsilon$, ${\rw'}^2(r_n)\rightarrow \infty$.  Also, $\Phi$ is bounded.  It follows that if $A{\rw'}^2$ is unbounded, then $\lim_{r\searrow r_0}A{\rw'}^2=\infty$.  However, Equation~(\ref{Aeq}) shows that this is impossible.  

Next, From Equation~(\ref{Neq}) it is clear that $\kappa\searrow \-\infty$ at least on a sequence as $\tau\nearrow \tau_0$.  As in Lemma~\ref{kinf}, $\tau_0<\infty$.  It follows that $\rw$ has a limit, $\rw_0$, as $r\searrow r_0$.

Equation~(\ref{feq}) establishes a limit $f_0$ for $A{\rw'}^2$.  To prove that $f_0=0$, we recall Equation~(\ref{veq})
\[r^2(A\rw')'+2r{\rw'}^2(A\rw')+\rw(1-\rw^2)=0.\]
If $f_0>0$, then $A{\rw'}^3\rightarrow \pm \infty$.  Consequently, $A\rw'$ has a nonzero limit which implies that $f_0=\infty$.  However, we have already proved that this is impossible. \hfill $\blacksquare$\\

We also have the following:
\begin{Crly}\label{wwpcurve}
For any black hole solution, $r_0\Phi_0\rw'_0+\rw_0(1-\rw_0^2)=0$.  In particular, $\lim_{r\searrow r_0}\rw'(r)$ exists and is finite.
\end{Crly}
\noindent
\textbf{Proof}:   Equation~(\ref{weq}) gives $\lim_{r\searrow r_0}\rw'(r)=\infty$ whenever $\lp_{r\searrow r_0}\rw'(r)=\infty$.  Equation~(\ref{feq}) then gives $f_0>0$, contradicting Lemma~\ref{bhkinf}.  Thus, $\lp_{r\searrow r_0}{\rw'}^2(r)<\infty$.  From Equation~(\ref{Aeq}), $A'$ has a finite limit as $r\searrow r_0$.  Solving Equation~(\ref{weq}) for $\rw''$ and integrating gives infinite $\rw'_0$ unless $r_0\Phi_0\rw'_0+\rw_0(1-\rw_0^2)=0$. \hfill $\blacksquare$\\   
As a consequence of Corollary~\ref{wwpcurve}, any black hole solution which is not a Reissner-Nordstr\"om solution, Equation~(\ref{ERN}), can be extended by a Kruskal-like change of coordinates beyond $r_0$ \cite{jS97}.

The uniqueness of the Reissner-Nordstr\"om solutions is established in the following:

\begin{Lma}\label{Ablowsup}
Suppose for some solution of Equations~(\ref{Aeq}) and~(\ref{weq}), there exists an $r_0$ such that $\lim_{r\searrow r_0}A(r)=\lim_{r\searrow r_0}\Phi(r)=\lim_{r\searrow r_0}f(r)=0$.  Then $\Lambda<1/4$ and the solution can be extended to $r_0=0$ and is a Reissner Nordstr\"om solution~(\ref{ERN}).
\end{Lma}
\noindent
\textbf{Proof}:  If $\rw_0(1-\rw_0)^2\ne 0$, then Equation~(\ref{weq}) implies $\rw'$ has a limit $\rw'_0$ as $r\searrow r_0$.  From Equation~(\ref{veq}), if $\rw_0<0$, then $\rw'_0\ge 0$.   If $\rw'_0=\infty$, then differentiating $\Phi$ gives $\Phi<0$ near $r_0$, which is impossible.  If, however, $\rw'_0<\infty$, then we make use of Equation~(\ref{Aeq}) which  gives $A<\epsilon(r-r_0)$ for any positive $\epsilon$, provided $r$ is sufficiently close to $r_0$.  Substituting this into Equation~(\ref{weq}) and integrating gives infinite $\rw'_0$ which is impossible.

Furthermore, $\rw_0^2\ne 1$.  Indeed, If $\rw_0=1$ and $\lp_{r\searrow r_0}{\rw'}^2(r)<\infty$, then $A''(r_0)<0$.  On the other hand, Equation~(\ref{weq}) implies that $\rw'$ can have only one sign near $r_0$.  Differentiating $\Phi$ again gives $\rw\rw'(1-\rw^2)>0$ which cannot occur.  This establishes that $\rw_0=0$.

Next, we consider the following two possibilities:\\ \\
(1) $r_0\ne \sqrt{2}$ and\\
(2) $r_0=\sqrt{2}$.\\ \\
\textit{Case} 1.  Equation~(\ref{zeq}) gives $z$ has one sign near $r_0$ and clearly, this sign must be positive.  We now consider the following variable:
\begin{equation}\label{h}
h(r)=A{\rw'}^2-\frac{\rw^2(\rw^2-2)}{2r^2}.
\end{equation}
A routine calculation yields
\begin{equation}\label{h'}
h'(r)=-{\rw'}^2z+\frac{\rw^2(\rw^2-2)}{r^3}.
\end{equation}   
From Equation~(\ref{h}) it is clear that $h(r)\ge 0$ for $r>r_0$.  However, it is also clear from Equation~(\ref{h'}) that $h'(r)\le 0$ in the same region.  Therefore, $\rw\equiv 0$.  Substituting this into Equation~(\ref{Aeq}) and integrating yields Equation~(\ref{ERN}).  $\Phi_0=\rw_0=0$ implies that $1-1/r_0-\Lambda r_0^2=0$ which can only occur if $\Lambda\le 1/4$.\\ \\
\textit{Case} 2.  We will prove that this case cannot occur.  Indeed, $r_0=\sqrt{2}$ only if $\Lambda=1/2$.  If there is such a solution, we choose some point $\tilde r$ near $\sqrt{2}$ and consider the equivalent solution of Equations~(\ref{req})~-~(\ref{keq}); i.e., the solution that satisfies $r(\tilde\tau)=\tilde r$, $N(\tilde\tau)=+\sqrt{A(\tilde r)}$, $\rw(\tilde\tau)=\rw(\tilde r)$, $U(\tilde\tau)=+\sqrt{A}\rw'(\tilde r)$, and $\kappa(\tilde\tau)=(\tilde r z(\tilde r)+2A(\tilde r))/\sqrt{4A}$ where $\tilde\tau$ is arbitrary.  By choosing $N$ positive, $\tau$ decreases as $r$ decreases to $\sqrt{2}$. 

We first prove that this solution can be continued back to $-\infty$ and $\lim_{\tau\searrow -\infty}\kappa(\tau)=0$.  Indeed, we define $\tau_0=\tau(\sqrt{2})$.  From Equation~(\ref{keq}) it follows immediately that $\lf_{\tau\searrow \tau_0}\kappa(\tau)>-\infty$.  If there exists a sequence $\{\tau_n\}\searrow \tau_0$ such that $\kappa(\tau_n)\rightarrow+\infty$, then Equation~(\ref{keq}) also gives $\lim_{\tau\searrow \tau_0}\kappa(\tau)=+\infty$.  In this case, $z>0$ for all $r\in(\sqrt{2},\tilde r)$ and, as in Case 1, Equation~(\ref{h}) gives $\rw\equiv 0$.  However, this is impossible.  Consequently, $\kappa$ is bounded; i.e., Equations~(\ref{req})~-~(\ref{keq}) are nonsingular at $\tau_0$.  Differentiating Equations~(\ref{req})~-~(\ref{keq}) indefinitely gives $\tau_c=-\infty$.  Equation~(\ref{keq}) now gives $\kappa\rightarrow 0$.
Because $-\kappa^2+\kappa N-N^2\le 0$, Equation~(\ref{k+N}) gives
\[\dot\kappa+\dot N<0.\]
Finally, $\kappa_0=N_0=0$, and $N>0$ for all $\tau\in(-\infty,\tilde\tau)$, yield
\begin{equation}\label{kappaneg}
\kappa(\tau)<0\;\;\;\mathrm{for\;all}\;\tau\in(-\infty,\tilde\tau).
\end{equation}
However, for at least some $\tau<\tilde\tau$, $A'$, and consequently, $\Phi$ and $\kappa$ must be positive.  \hfill $\blacksquare$

\bibliographystyle{plain}
\bibliography{bgen}

%\nocite{nS96}
%\nocite{pB94}
%\nocite{wR77}
\end{document}